\documentclass[aps,prb,reprint,showpacs,floatfix]{revtex4-1}
\usepackage{amsmath,graphicx,color,verbatim}

\renewcommand{\vec}[1]{\mathbf{#1}}

\begin{document}

\title{Remarkable thermoelectric performance in BaPdS$_2$ via pudding-mold band structure and ultralow lattice thermal conductivity}
\author{Eric B. Isaacs}
\author{Chris Wolverton}
\email{c-wolverton@northwestern.edu}
\affiliation{Department of Materials Science and Engineering, Northwestern University, Evanston, Illinois 60208, USA}

\begin{abstract}
Efficient thermoelectric materials require a rare and contraindicated
combination of materials properties: large electrical conductivity,
large Seebeck coefficient, and low thermal conductivity. One strategy
to achieve the first two properties is via low-energy electronic bands
containing both flat and dispersive parts in different regions of
crystal momentum space, known as a pudding-mold band structure. Here,
we illustrate that BaPdS$_2$ successfully achieves the pudding-mold
band structure, contributing to a large thermoelectric power factor,
due to its anisotropic crystal structure containing zig-zag chains of
edge-sharing square planar PdS$_4$ units. In addition, BaPdS$_2$
exhibits ultralow lattice thermal conductivity, and thus also achieves
the third property, due to extremely soft and anharmonic interactions
in its transverse acoustic phonon branch. We predict a remarkably
large thermoelectric figure of merit, with peak values between 2 and 3
for two of the three crystallographic directions, suggesting BaPdS$_2$
warrants experimental investigation.
\end{abstract}

\date{\today}
\maketitle

\section{Introduction}\label{sec:intro}

The figure of merit for a thermoelectric material, which can convert a
temperature gradient into electrical current, is
\begin{equation}ZT =
\frac{S^2\sigma }{\kappa_{el} + \kappa_L}T,\label{eq:zt}\end{equation}
where $S$, $\sigma$, $T$, $\kappa_{el}$, and $\kappa_L$ are the
Seebeck coefficient, electrical conductivity, temperature, electrical
contribution to thermal conductivity, and lattice contribution to
thermal conductivity,
respectively.\cite{goldsmid_introduction_2010,zevalkink_practical_2018}
One strategy to achieve a large $ZT$ is via engineering the electronic
band structure, i.e., the electron energy $\varepsilon$ as a function
of crystal momentum $\vec{k}$.\cite{pei_band_2012} In particular, a
band with both flat (small $\nabla_{\vec{k}}\varepsilon$) and
dispersive (large $\nabla_{\vec{k}}\varepsilon$) parts along a
direction in $\vec{k}$-space has been shown to enhance $\sigma S^2$ if
the electron chemical potential $\mu$ lies at an energy separating the
two regions.\cite{kuroki_pudding_2007} In such a ``pudding-mold'' band
structure, the band velocity difference leads to large $S$, and large
$\sigma$ is achieved due to the dispersive part of the band and a
large Fermi surface.\cite{kuroki_pudding_2007} More generally, one can
consider a broader definition of a pudding-mold band structure in
which the flat and dispersive regions can along different directions
in $\vec{k}$-space, which naturally occurs for low-dimensional bulk
and nanostructured
crystals.\cite{hicks_effect_1993,hicks_thermoelectric_1993,usui_enhanced_2017}
The pudding-mold band structure has been used to explain the promising
thermoelectric performance of Na$_x$CoO$_2$,\cite{kuroki_pudding_2007}
SnSe,\cite{zhao_ultralow_2014,kutorasinski_electronic_2015,wang_defects_2018}
and recently-proposed Fe-based Heusler
compounds\cite{bilc_low-dimensional_2015} among others.

Recently, the promising thermoelectric performance of several Pd oxide
compounds has been attributed in part to the pudding-mold band
structure. The layered compound PbPdO$_2$, for example, exhibits a
large Seebeck coefficient of 175 $\mu$V/K at 600 K when hole
doped.\cite{lamontagne_high_2016} Based on first-principles
calculations, hole-doped Bi$_2$PdO$_4$ was predicted to exhibit high
power factor ($\sigma S^2$) in addition to low
$\kappa_L$.\cite{he_bi2pdo4_2017} In both of these compounds,
achieving the pudding-mold band structure appears to be connected to
the presence of square planar Pd$^{2+}$. Based on this observation, we
previously developed an inverse band structure design approach based
on a materials database screening to search for other square planar
compounds achieving the pudding-mold band
structure.\cite{isaacs_inverse_2018} Although several chemistries were
considered, we focused on oxide materials such as Ba$_2$PdO$_3$ and
La$_4$PdO$_7$ in order to validate the approach.

The thermoelectric efficiency of oxides is often limited by (1) large
$\kappa_L$, stemming from the low atomic mass of oxygen and rigid
chemical bonding, and (2) low carrier
mobility.\cite{he_oxide_2011,garrity_first-principles_2016} Therefore,
in this work, we turn our attention to \textit{chalcogenide}
materials. Chalcogenides comprise many of the most-investigated and
highest-performance thermoelectrics, such as PbTe, Bi$_2$Te$_3$, SnSe,
Cu$_2$(S/Se), and
La$_{3-x}$Te$_4$.\cite{snyder_complex_2008,zhang_thermoelectric_2015,tan_rationally_2016}
We investigate a chalcogenide material based on square planar Pd,
namely BaPdS$_2$, with the aim of achieving low $\kappa_L$ in addition
to pudding-mold band structure. BaPdS$_2$ is the only such square
planar chalcogenide identified by our previous inverse band structure
design approach\cite{isaacs_inverse_2018} other than metallic
EuPd$_3$S$_4$\cite{yao_first_2004} and the binaries
PdS$_2$\cite{bhatt_growth_2013} and
PdSe$_2$.\cite{sun_electronic_2015} Although BaPdS$_2$ was first
synthesized in 1986,\cite{huster_synthesis_1986} it has not been
explored as a thermoelectric material to the best of our knowledge.
The presence of heavy Ba, in addition to the lack of oxygen, leads to
the potential for low $\kappa_L$. Using electronic structure and
transport calculations based on the density functional theory, we find
that BaPdS$_2$ exhibits (1) a pudding-mold band structure for $p$-type
doping, (2) large power factor for both $n$- and $p$-type doping, (3)
ultralow $\kappa_L$ due to extremely soft and anharmonic bonding, and
(4) highly-anisotropic electronic and thermal transport behavior. We
find remarkable predicted thermoelectric performance, with $ZT$ larger
than 2 in certain crystallographic directions at high temperature,
assuming optimal doping and using a reasonable value of electronic
relaxation time $\tau$ (5
fs).\cite{madsen_automated_2006,gandi_ws2_2014,bilc_low-dimensional_2015,wang_thermoelectric_2015,hao_computational_2016,pal_bonding_2018}
Based on these results, we suggest BaPdS$_2$ warrants experimental
investigation.

\section{Computational Details}\label{sec:compdetails}

We perform density functional
theory\cite{hohenberg_inhomogeneous_1964,kohn_self-consistent_1965}
calculations using the generalized gradient
approximation\cite{perdew_generalized_1996} and the projector
augmented wave method (Ba\_sv, Pd, and S potentials)
\cite{blochl_projector_1994,kresse_ultrasoft_1999} in the Vienna
\textit{ab initio} simulation package
(\textsc{vasp}).\cite{kresse_ab_1994,kresse_ab_1993,kresse_efficient_1996,kresse_efficiency_1996}
A plane wave basis with 500 eV kinetic energy cutoff and uniform
$k$-meshes with $\ge 700$ $k$-points per \AA$^{-3}$ are employed. The
ionic forces and total energy are converged to 10$^{-3}$ eV/\AA\ and
10$^{-6}$ eV, respectively. High-symmetry $k$- and $q$-point paths are
based on the conventions of Setyawan \textit{et
  al.}\cite{setyawan_high-throughput_2010}

Semiclassical electronic transport calculations within the constant
relaxation time ($\tau$) approximation are performed using
\textsc{boltztrap}\cite{madsen_boltztrap_2006} with $k$-point density
of 3,500/\AA$^{-3}$. Phonon calculations are performed using the
direct approach in \textsc{phonopy}\cite{phonopy} with a 192-atom
supercell chosen to be approximately cubic based on the algorithm of
Erhart \textit{et al.}\cite{erhart_first-principles_2015} as
implemented in \textsc{ase}.\cite{ase-paper} The mode Gr\"{u}neisen
parameter is computed using a $\pm 0.3\%$ volume difference.
$\kappa_L$ is computed via the Debye-Callaway
model\cite{callaway_model_1959,morelli_estimation_2002,zhang_first-principles_2016}
and compared to that of the minimum thermal conductivity model of
Cahill \textit{et al.}\cite{cahill_lower_1992} $\kappa_{el}$ is
computed via the Wiedemann-Franz law. Additional details on the
electronic and thermal transport calculations are included in the
Supplemental Material.\footnote{See Supplemental Material for
  additional details on electronic and thermal transport methods,
  phonon and anharmonicity calculation convergence, animations of
  several low-frequency phonon modes, comparison to experimental
  crystal structure and formation energy, and the electronic band
  structure for a different high-symmetry crystal momentum path.}

\section{Results and Discussion}\label{sec:results}

\begin{figure}[htbp]
  \begin{center}
    \includegraphics[width=1.0\linewidth]{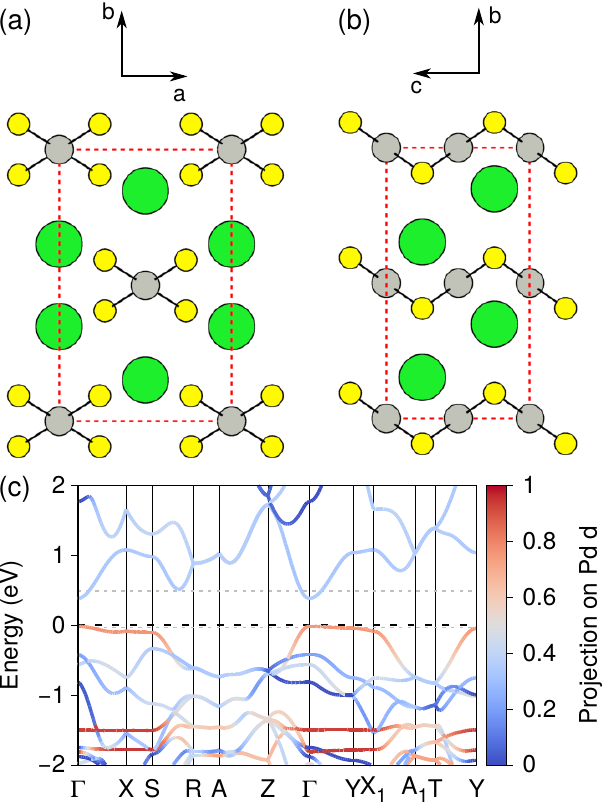}
  \end{center}
  \caption{Orthographic projection of the crystal structure of
    BaPdS$_2$ along the (a) \textbf{c} and (b) \textbf{a} axes of the
    conventional unit cell (red dashed lines). The green, gray, and
    yellow circles represent Ba, Pd, and S atoms, respectively. Black
    lines indicate the Pd--S bonds of the square planar units. (c) The
    electronic bands of BaPdS$_2$ colored by the Pd $d$ character of
    the electronic states. The black dashed line corresponds to the
    valence band minimum, and the gray horizontal dotted lines
    indicate the $\pm 10^{20}$ cm$^{-3}$ doping levels.}
  \label{fig:structure_bands}
\end{figure}

BaPdS$_2$, shown in Fig. \ref{fig:structure_bands}(a) and Fig.
\ref{fig:structure_bands}(b), has a C-centered orthorhombic unit cell
with space group $Cmcm$. Isostructural to BaNiO$_2$, it consists of 1D
chains of edge-sharing PdS$_4$ square planar units (corresponding to a
PdS$_2$ composition) connected in a zig-zag pattern. The computed
lattice parameters, atomic positions, and formation energy are in good
agreement with experimental values, as discussed in the Supplemental
Material. The electronic band structure of BaPdS$_2$ is shown in Fig.
\ref{fig:structure_bands}(c). The valence band, which consists
primarily of Pd $d$ states, exhibits a pudding-mold band structure. It
is dispersive in the direction along the 1D chains (e.g., $\Gamma$--Z)
and flat along other directions (e.g., $\Gamma$--X and $\Gamma$--Y).
In order words, the pudding-mold band structure is closely related to
the highly-anisotropic crystal structure of BaPdS$_2$. We do not find
a pudding-mold band structure for the conduction band, which consists
of a hybridization of Pd $d$ and S $p$ orbitals. We note that an
additional conduction band minimum between S and R is close in energy
($\sim$130 meV) to the band edge at $\Gamma$.

\begin{figure}[htbp]
  \begin{center}
    \includegraphics[width=1.0\linewidth]{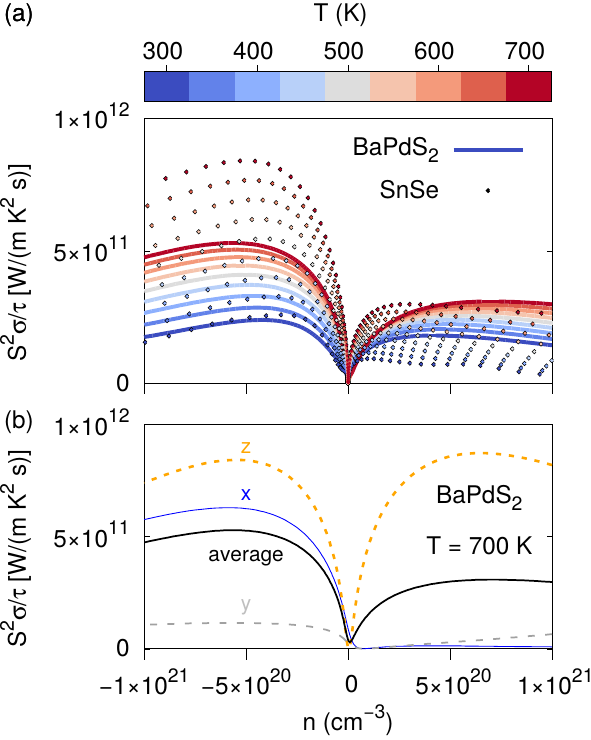}
  \end{center}
  \caption{(a) Power factor divided by electronic relaxation time as a
    function of carrier concentration for BaPdS$_2$ (lines) and SnSe
    (circles), averaged over the $x$, $y$, and $z$ directions, for
    several values of $T$. (b) $T=700$ K power factor divided by
    electronic relaxation time versus carrier concentration in each
    direction, in addition to the average, for BaPdS$_2$.}
  \label{fig:power_factor}
\end{figure}

The band structure of BaPdS$_2$ leads to notable electronic transport
properties. Figure \ref{fig:power_factor}(a) illustrates the behavior
of the average (over Cartesian directions) of $\sigma S^2/\tau$ as a
function of carrier concentration for several temperatures. Results
for BaPdS$_2$ are shown in comparison to those of SnSe, a
high-performing thermoelectric
material.\cite{zhao_ultralow_2014,zhao_ultrahigh_2015,duong_achieving_2016,chang_3d_2018}
BaPdS$_2$ achieves comparable $p$-type power factor behavior to SnSe,
albeit at larger (though still reasonable\cite{snyder_complex_2008})
doping values. For example, at 700 K, the maximum predicted $p$-type
$\sigma S^2/\tau$ is $\sim 3 \times 10^{11}$ W/(m$\cdot$K$^2\cdot$s)
for both BaPdS$_2$ (for $7 \times 10^{20}$ holes/cm$^3$) and SnSe (for
$3 \times 10^{20}$ holes/cm$^3$). We also note that the $p$-type power
factor behavior exhibits significantly less temperature dependence for
BaPdS$_2$ than for SnSe. Although the conduction band does not exhibit
pudding-mold qualities, we find even larger $\sigma S^2/\tau$ for
$n$-type doping, though smaller than the corresponding SnSe behavior.
For example, the peak $n$-type $\sigma S^2/\tau$ for BaPdS$_2$ at 700
K is $5.3 \times 10^{11}$ W/(m$\cdot$K$^2\cdot$s), occurring at a
doping value of $\sim 5.5 \times 10^{20}$ electrons/cm$^3$; SnSe
exhibits a significantly larger peak $n$-type $\sigma S^2/\tau$ of
$8.4 \times 10^{11}$ W/(m$\cdot$K$^2\cdot$s) at around the same doping
magnitude. The appreciable $n$-type power factor behavior for
BaPdS$_2$ likely stems from the band
convergence,\cite{pei_convergence_2011,liu_convergence_2012,tang_convergence_2015,zeier_thinking_2016,zeier_new_2017},
i.e., the small energy separation between the conduction band minima
(1) at $\Gamma$ and (2) between S and R.

Due to the structural anisotropy, the power factor behavior for
BaPdS$_2$ is highly anisotropic. We illustrate the power factor
behavior at 700 K, for example, in Fig. \ref{fig:power_factor}(b) for
each direction. For hole doping, $\sigma S^2/\tau$ is only appreciable
along the $z$ direction (i.e., along the 1D chains), which is
consistent with the pudding-mold band structure. In contrast, for
electron doping, large $\sigma S^2/\tau$ is observed in both the $z$
and $x$ directions. A rationalization for the lower $n$-type power
factor behavior in the $y$ direction is discussed in the Supporting
Information.

\begin{figure}[htbp]
  \begin{center}
    \includegraphics[width=1.0\linewidth]{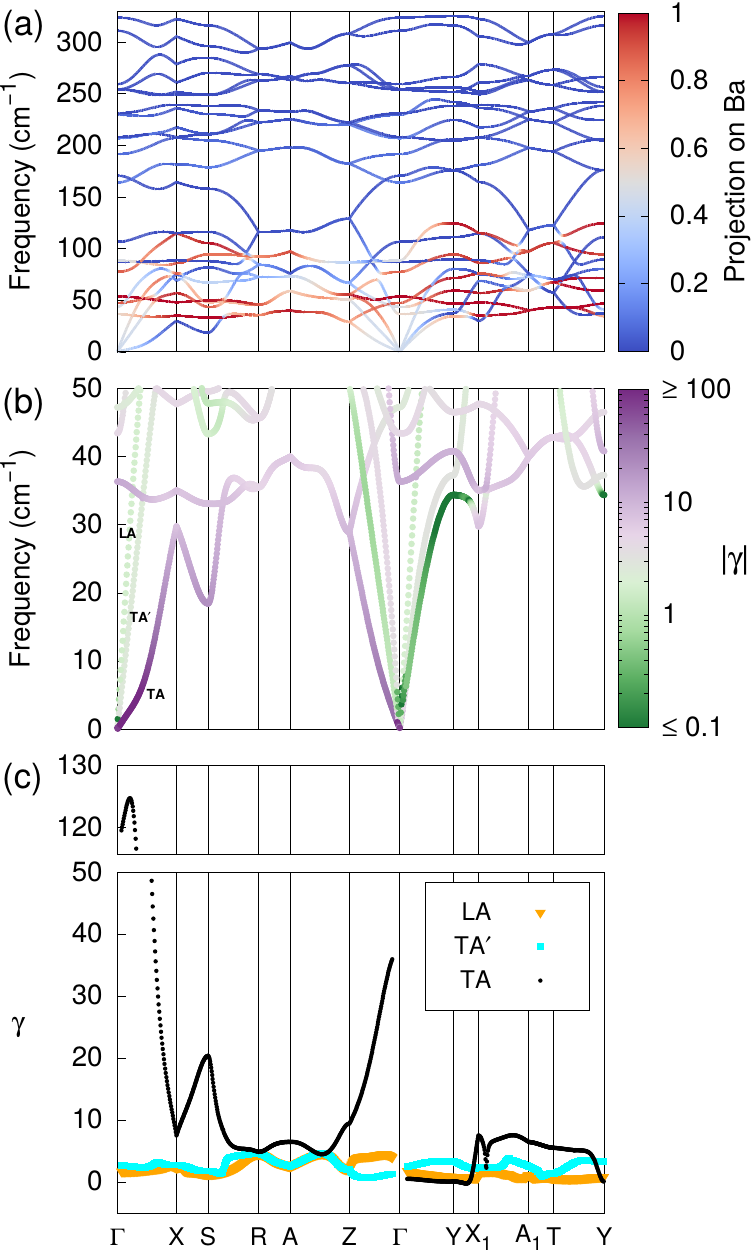}
  \end{center}
  \caption{(a) BaPdS$_2$ phonon bands colored by Ba contribution to
    the phonon eigenvector. (b) The low-frequency portion of the
    phonon dispersion with bands colored by the magnitude of the mode
    Gr\"{u}neisen parameter. (c) Dispersion of the mode Gr\"{u}neisen
    parameter $\gamma_{\vec{q}\nu}$ for the acoustic branches. Due to
    the large $\gamma_{\vec{q}\nu}$ values for the TA branch along
    $\Gamma$--X, we use a broken $\gamma_{\vec{q}\nu}$ axis for
    clarity.}
  \label{fig:phonons}
\end{figure}

\begin{table}[htbp]
  \renewcommand{\arraystretch}{1.2}
  \begin{tabular}{|c|c|c|c|c|c|c|}\hline
    Branch & $v_x$ & $v_y$ & $v_z$ & $\Theta_{\Gamma X}$ & $\Theta_{\Gamma Y}$ & $\Theta_{\Gamma Z}$\\ \hline
    TA & 0.5 & 1.9 & 0.8 & 43 & 49 & 42\\
    TA$'$ & 2.4 & 2.4 &1.9 & 82 & 54 & 72\\ 
    LA & 4.1 & 3.7 & 3.6 & 99 & 115 & 72\\\hline
  \end{tabular}
  \caption{Group velocity $v$ at $\Gamma$ (in km/s) and Debye
    temperature $\Theta$ (in K) of BaPdS$_2$ for each acoustic branch
    and direction.}
  \label{tab:harmonic}
\end{table}


Having established the promising predicted electronic transport
properties of BaPdS$_2$, we now turn our attention to thermal
transport. The phonon dispersion of BaPdS$_2$ is shown in Fig.
\ref{fig:phonons}(a). Low-frequency optical modes, primarily involving
vibration of the heavy Ba atoms, are observed in the $\sim$ 30--110
cm$^{-1}$ frequency range. For each acoustic branch, we compute the
group velocity in the long-wavelength limit ($v_x$, $v_y$, and $v_z$)
and Debye temperature ($\Theta_{\Gamma X}$, $\Theta_{\Gamma Y}$, and
$\Theta_{\Gamma Z}$) for each direction, as shown in Table
\ref{tab:harmonic}. Here, we define $\Theta$ as the acoustic phonon
frequency at the zone boundary, and the acoustic branch definitions
are given in the Supplemental Material. BaPdS$_2$ is a very soft
material, exhibiting low group velocities (0.5--4.1 km/s) and Debye
temperatures (42--115 K). For comparison, $v$ of 1.0--2.9 and $\Theta$
of 19--72 are found in SnSe.\cite{zhao_ultralow_2014} The transverse
acoustic (TA) branch of BaPdS$_2$ is particular soft, with $v=0.5$
km/s ($0.8$ km/s) and $\Theta=43$ K ($42$ K) along the $x$ ($z$)
direction. Animations of the TA phonons along X (featuring rigid
PdS$_2$ chains sliding in the $z$ direction with respect to the
others) and along Z (featuring oscillations of the individual PdS$_2$
chains in the $x$ direction) are included in the Supplemental
Material. We note that the behavior of the TA branch is significantly
harder along the $y$ direction in terms of the group velocity ($1.9$
km/s), though not the Debye temperature ($49$ K). Corresponding to
PdS$_2$ chains sliding in the $z$ direction, the TA branch along Y
corresponds to a similar atomic motion to that along X; the
significantly larger inter-chain distance in the $y$ direction (10.8
\AA) compared to that in $x$ direction (6.9 \AA) may be responsible
for the very different phonon properties.




\begin{table}[htbp]
  \renewcommand{\arraystretch}{1.2}
  \begin{tabular}{|c|c|c|c|c|c|c|c|}\cline{2-8}
    \multicolumn{1}{c}{} & \multicolumn{4}{|c|}{BaPdS$_2$} & \multicolumn{3}{|c|}{SnSe} \\\hline
    Branch & $\gamma_{all}$ & $\gamma_{\Gamma X}$ & $\gamma_{\Gamma Y}$ & $\gamma_{\Gamma Z}$ & $\gamma_{\Gamma X}$ & $\gamma_{\Gamma Y}$ & $\gamma_{\Gamma Z}$\\ \hline
    TA    & 25.5  & 80.3 & 0.3 & 22.8 & 5.1 & 2.7 & 1.0\\
    TA$'$ &  2.8  & 2.6  & 3.1 & 1.0  & 1.1 & 2.4 & 2.7 \\ 
    LA    &  2.3  & 2.1  & 1.2 & 4.0  & 5.9 & 1.3 & 3.3 \\\hline
    Avg.  & 10.2  & 28.3 & 1.5 & 9.3  & 4.1 & 2.1 & 2.3 \\\hline
  \end{tabular}
  \caption{Averages (computed as $\sqrt{\langle\gamma^2\rangle}$) of
    mode Gr\"{u}neisen parameter $\gamma_{\vec{q}\nu}$ for BaPdS$_2$
    and SnSe\cite{zhao_ultralow_2014} for different acoustic branches
    and directions. For BaPdS$_2$, the average over the entire
    high-symmetry path in the Brillouin zone is also given as $\gamma_{all}$.
  }
  \label{tab:anharmonic}
\end{table}


To acquire a baseline assessment for the overall magnitude of the
anharmonicity, i.e., the phonon-phonon scattering strength, we compute
the mode Gr\"{u}neisen parameter \begin{equation}\gamma_{\vec{q}\nu}
  =-\frac{\partial\omega_{\vec{q}\nu}/\omega_{\vec{q}\nu}}{\partial
    V/V},\end{equation} where $V$ is the volume and
$\omega_{\vec{q}\nu}$ is the phonon frequency for crystal momentum
$\vec{q}$ and branch $\nu$. The low-frequency phonons of BaPdS$_2$,
colored by $|\gamma|$, are shown in Fig. \ref{fig:phonons}(b), whereas
the full dispersion of $\gamma_{\vec{q}\nu}$ for the acoustic branches
is included in Fig. \ref{fig:phonons}(c). The most prominent feature
is that BaPdS$_2$ has gigantic $\gamma_{\vec{q}\nu}$ for the TA branch
along $\Gamma$--X (values up to $\sim125$) and $\Gamma$--Z (values up
to $\sim35$), which indicates extremely anharmonic interactions for
this acoustic branch in BaPdS$_2$. We perform tests to confirm the
large computed $\gamma_{\vec{q}\nu}$ are not a numerical artifact, as
discussed in the Supplemental Material. Averages of the mode
Gr\"{u}neisen parameter for each acoustic branch of BaPdS$_2$,
averaged over $\Gamma$--X, $\Gamma$--Y, $\Gamma$--Z, and the full
high-symmetry path in the Brillouin zone, are shown in Table
\ref{tab:anharmonic}, in which we also compare to literature results
for SnSe.\cite{zhao_ultralow_2014} The average $\gamma_{\vec{q}\nu}$
for the TA branch of BaPdS$_2$ is 80.3 in the $x$ direction and 22.8
in the $z$ direction, compared to much smaller (less anharmonic)
values of 5.1 and 1.0 for SnSe in the corresponding directions. We
note that a key exception to the large anharmonicity of the TA branch
of BaPdS$_2$ is along the $y$ direction, for which the average
$\gamma_{\vec{q}\nu}$ is only 0.3, as compared to 2.7 for SnSe. In
other words, the TA branch of BaPdS$_2$ is not only harder in the $y$
direction than the other directions, but also much less anharmonic.
Although the most remarkable finding is the extremely large
$\gamma_{\vec{q}\nu}$ for the TA branch, we also find significant
anharmonicity for other phonon branches. For example, the TA$'$ (LA)
branch exhibits appreciable $\gamma_{\vec{q}\nu}$ of 1.0--3.1
(1.2--4.0), as compared to 1.1--2.7 (1.3--5.9) for SnSe. Large
anharmonicity is also found in the lowest-frequency optical branch,
with mode Gr\"{u}neisen parameter values as large as $\sim 12$, as
shown in Fig. \ref{fig:phonons}(b). This branch corresponds to
oscillation of the Ba sublattice with respect to the PdS$_2$ chains in
the $z$ direction.


\begin{figure*}[htbp]
  \begin{center}
    \includegraphics[width=1.0\linewidth]{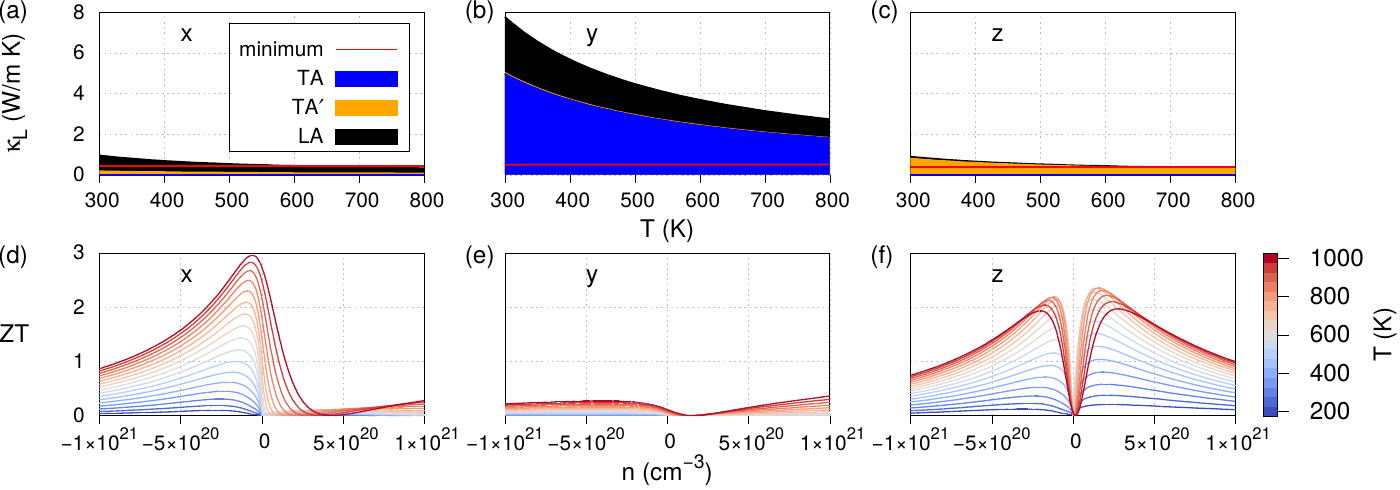}
  \end{center}
  \caption{(a)--(c) Stacked plots of BaPdS$_2$ acoustic branch
    contributions to $\kappa_L$ versus $T$ computed via the
    Debye-Callaway model for the (a) $x$, (b) $y$, and (c) $z$
    directions, respectively. Red lines indicate the minimum
    $\kappa_L$ of the Cahill model. (d)--(f) $ZT$ as a function of
    carrier concentration for different $T$ in the (d) $x$, (e) $y$,
    and (f) $z$ directions, respectively. Values are shown for
    $\tau=5$ fs.}
  \label{fig:kappa_zt}
\end{figure*}

We employ the Debye-Callaway model, an approximate model taking into
account normal and Umklapp acoustic phonon-phonon scattering via the
mode Gr\"{u}neisen parameter, as a means to estimate the lattice
thermal conductivity of
BaPdS$_2$.\cite{callaway_model_1959,morelli_estimation_2002,zhang_first-principles_2016}
We note that the low-frequency optical phonons with large
$\gamma_{\vec{q}\nu}$, not taken into account in the Debye-Callaway
model, may further lower $\kappa_L$ compared to the values reported
here. The contributions of each acoustic phonon branch to $\kappa_L$
as a function of $T$ in the $x$, $y$, and $z$ directions are shown in
Fig. \ref{fig:kappa_zt}(a)--\ref{fig:kappa_zt}(c), respectively.
BaPdS$_2$ exhibits ultralow lattice thermal conductivity ($\kappa_L$
less than 1 W/m$\cdot$K) in the $x$ and $z$ directions. In these
directions, the predicted $\kappa_L$ are of the same magnitude as that
of SnSe along its smallest-$\kappa_L$ direction (values of $\sim$
0.25--0.7 W/m$\cdot$K).\cite{zhao_ultralow_2014} For BaPdS$_2$, the
main contributor to $\kappa_L$ in the $x$ direction is the LA branch,
which exhibits the largest group velocity; the main contributor in the
$z$ direction is the TA$'$ branch, which is the least anharmonic. Like
the power factor, the thermal transport is highly anisotropic: in the
$y$ direction, $\kappa_L$ is much larger (values of 4--8 W/m$\cdot$K)
than in the other directions, stemming primarily from the small
$\gamma_{\Gamma Y}$ of 0.3 for the TA branch, as discussed above. In
the $x$ and $z$ (but not $y$) directions, BaPdS$_2$ achieves the
minimum possible $\kappa_L$, shown as red lines in Fig.
\ref{fig:kappa_zt}(a)--\ref{fig:kappa_zt}(c), estimated from the model
of Cahill \textit{et al.} Overall, BaPdS$_2$ is predicted to exhibit
remarkably poor thermal transport in directions other than $y$.

As illustrated in Fig. \ref{fig:kappa_zt}(d)--\ref{fig:kappa_zt}(f),
the combination of large power factor and low $\kappa_L$ leads to
impressive figure of merit for BaPdS$_2$, particularly in the $x$
direction for electron doping and in the $z$ direction for both
electron and hole doping. At high temperature, we find peak $ZT$
values of nearly 3 in the $x$ direction (for $6.0\times 10^{19}$
electrons/cm$^3$) and values of nearly 2.3 in the $z$ direction (for
$1.1\times 10^{20}$ electrons/cm$^3$ or $1.6\times 10^{20}$
holes/cm$^3$). In contrast, much smaller peak $ZT$ ($<1$) is observed
for BaPdS$_2$ in the $y$ direction, due to the smaller power factor
and larger $\kappa_L$. In order to obtain the predicted $ZT$ values
discussed above, we have chosen an electronic relaxation time ($\tau$)
of 5 fs. While an \textit{ab initio} estimation of $\tau$ (e.g., from
electron-phonon scattering calculations) will be important future
work, we note that 5 fs is reasonable in terms of the order of
magnitude\cite{ashcroft_mermin_1976,yu_cardona_1996} and can be
considered a conservative estimate based on previous
works.\cite{madsen_automated_2006,gandi_ws2_2014,bilc_low-dimensional_2015,wang_thermoelectric_2015,hao_computational_2016,pal_bonding_2018}
Even with a less optimistic guess for the electronic scattering time,
we still find quite large values for the thermoelectric figure of
merit, albeit at larger doping. For example, assuming $\tau=1$ fs, a
peak $ZT$ of 1.3 is achieved in the $x$ direction for $2.0\times
10^{20}$ electrons/cm$^3$. Therefore, even with the uncertainty in
$\tau$, our results strongly suggest BaPdS$_2$ merits experimental
investigation.

\section{Conclusions}\label{sec:conclusions}

BaPdS$_2$, previously identified by our inverse band structure design
approach, is a square planar sulfide material with remarkable,
anisotropic thermoelectric properties. Due to a pudding-mold valence
band structure and multiple conduction bands, BaPdS$_2$ exhibits
impressive power factor behavior in two of the crystallographic
directions. With heavy Ba atoms and extremely soft and anharmonic
bonding, BaPdS$_2$ achieves ultralow lattice thermal conductivity in
the $x$ and $z$ directions. We predict peak $ZT$ values of 2--3 in the
$x$ direction ($n$-type) and the $z$ direction ($n$- and $p$-type) for
an electronic relaxation time of 5 fs. Our results strongly suggest
BaPdS$_2$ warrants experimental investigation for its remarkable
electronic, thermal transport, and thermoelectric properties.

\begin{acknowledgments}
We acknowledge support from the U.S. Department of Energy under
Contract DE-SC0014520. Computational resources were provided by the
National Energy Research Scientific Computing Center (U.S. Department
of Energy Contract DE-AC02-05CH11231) and the Extreme Science and
Engineering Discovery Environment (National Science Foundation
Contract ACI-1548562).
\end{acknowledgments}

\bibliography{bapds2}

\begin{thebibliography}{57}%
\makeatletter
\providecommand \@ifxundefined [1]{%
 \@ifx{#1\undefined}
}%
\providecommand \@ifnum [1]{%
 \ifnum #1\expandafter \@firstoftwo
 \else \expandafter \@secondoftwo
 \fi
}%
\providecommand \@ifx [1]{%
 \ifx #1\expandafter \@firstoftwo
 \else \expandafter \@secondoftwo
 \fi
}%
\providecommand \natexlab [1]{#1}%
\providecommand \enquote  [1]{``#1''}%
\providecommand \bibnamefont  [1]{#1}%
\providecommand \bibfnamefont [1]{#1}%
\providecommand \citenamefont [1]{#1}%
\providecommand \href@noop [0]{\@secondoftwo}%
\providecommand \href [0]{\begingroup \@sanitize@url \@href}%
\providecommand \@href[1]{\@@startlink{#1}\@@href}%
\providecommand \@@href[1]{\endgroup#1\@@endlink}%
\providecommand \@sanitize@url [0]{\catcode `\\12\catcode `\$12\catcode
  `\&12\catcode `\#12\catcode `\^12\catcode `\_12\catcode `\%12\relax}%
\providecommand \@@startlink[1]{}%
\providecommand \@@endlink[0]{}%
\providecommand \url  [0]{\begingroup\@sanitize@url \@url }%
\providecommand \@url [1]{\endgroup\@href {#1}{\urlprefix }}%
\providecommand \urlprefix  [0]{URL }%
\providecommand \Eprint [0]{\href }%
\providecommand \doibase [0]{http://dx.doi.org/}%
\providecommand \selectlanguage [0]{\@gobble}%
\providecommand \bibinfo  [0]{\@secondoftwo}%
\providecommand \bibfield  [0]{\@secondoftwo}%
\providecommand \translation [1]{[#1]}%
\providecommand \BibitemOpen [0]{}%
\providecommand \bibitemStop [0]{}%
\providecommand \bibitemNoStop [0]{.\EOS\space}%
\providecommand \EOS [0]{\spacefactor3000\relax}%
\providecommand \BibitemShut  [1]{\csname bibitem#1\endcsname}%
\let\auto@bib@innerbib\@empty
\bibitem [{\citenamefont {Goldsmid}(2010)}]{goldsmid_introduction_2010}%
  \BibitemOpen
  \bibfield  {author} {\bibinfo {author} {\bibfnamefont {H.~J.}\ \bibnamefont
  {Goldsmid}},\ }\href {http://link.springer.com/10.1007/978-3-642-00716-3}
  {\emph {\bibinfo {title} {Introduction to {Thermoelectricity}}}},\ \bibinfo
  {series} {Springer {Series} in {Materials} {Science}}, Vol.\ \bibinfo
  {volume} {121}\ (\bibinfo  {publisher} {Springer},\ \bibinfo {address}
  {Berlin, Heidelberg},\ \bibinfo {year} {2010})\BibitemShut {NoStop}%
\bibitem [{\citenamefont {Zevalkink}\ \emph {et~al.}(2018)\citenamefont
  {Zevalkink}, \citenamefont {Smiadak}, \citenamefont {Blackburn},
  \citenamefont {Ferguson}, \citenamefont {Chabinyc}, \citenamefont {Delaire},
  \citenamefont {Wang}, \citenamefont {Kovnir}, \citenamefont {Martin},
  \citenamefont {Schelhas}, \citenamefont {Sparks}, \citenamefont {Kang},
  \citenamefont {Dylla}, \citenamefont {Snyder}, \citenamefont {Ortiz},\ and\
  \citenamefont {Toberer}}]{zevalkink_practical_2018}%
  \BibitemOpen
  \bibfield  {author} {\bibinfo {author} {\bibfnamefont {A.}~\bibnamefont
  {Zevalkink}}, \bibinfo {author} {\bibfnamefont {D.~M.}\ \bibnamefont
  {Smiadak}}, \bibinfo {author} {\bibfnamefont {J.~L.}\ \bibnamefont
  {Blackburn}}, \bibinfo {author} {\bibfnamefont {A.~J.}\ \bibnamefont
  {Ferguson}}, \bibinfo {author} {\bibfnamefont {M.~L.}\ \bibnamefont
  {Chabinyc}}, \bibinfo {author} {\bibfnamefont {O.}~\bibnamefont {Delaire}},
  \bibinfo {author} {\bibfnamefont {J.}~\bibnamefont {Wang}}, \bibinfo {author}
  {\bibfnamefont {K.}~\bibnamefont {Kovnir}}, \bibinfo {author} {\bibfnamefont
  {J.}~\bibnamefont {Martin}}, \bibinfo {author} {\bibfnamefont {L.~T.}\
  \bibnamefont {Schelhas}}, \bibinfo {author} {\bibfnamefont {T.~D.}\
  \bibnamefont {Sparks}}, \bibinfo {author} {\bibfnamefont {S.~D.}\
  \bibnamefont {Kang}}, \bibinfo {author} {\bibfnamefont {M.~T.}\ \bibnamefont
  {Dylla}}, \bibinfo {author} {\bibfnamefont {G.~J.}\ \bibnamefont {Snyder}},
  \bibinfo {author} {\bibfnamefont {B.~R.}\ \bibnamefont {Ortiz}}, \ and\
  \bibinfo {author} {\bibfnamefont {E.~S.}\ \bibnamefont {Toberer}},\ }\href
  {\doibase 10.1063/1.5021094} {\bibfield  {journal} {\bibinfo  {journal}
  {Appl. Phys. Rev.}\ }\textbf {\bibinfo {volume} {5}},\ \bibinfo {pages}
  {021303} (\bibinfo {year} {2018})}\BibitemShut {NoStop}%
\bibitem [{\citenamefont {Pei}\ \emph {et~al.}(2012)\citenamefont {Pei},
  \citenamefont {Wang},\ and\ \citenamefont {Snyder}}]{pei_band_2012}%
  \BibitemOpen
  \bibfield  {author} {\bibinfo {author} {\bibfnamefont {Y.}~\bibnamefont
  {Pei}}, \bibinfo {author} {\bibfnamefont {H.}~\bibnamefont {Wang}}, \ and\
  \bibinfo {author} {\bibfnamefont {G.~J.}\ \bibnamefont {Snyder}},\ }\href
  {\doibase 10.1002/adma.201202919} {\bibfield  {journal} {\bibinfo  {journal}
  {Adv. Mater.}\ }\textbf {\bibinfo {volume} {24}},\ \bibinfo {pages} {6125}
  (\bibinfo {year} {2012})}\BibitemShut {NoStop}%
\bibitem [{\citenamefont {Kuroki}\ and\ \citenamefont
  {Arita}(2007)}]{kuroki_pudding_2007}%
  \BibitemOpen
  \bibfield  {author} {\bibinfo {author} {\bibfnamefont {K.}~\bibnamefont
  {Kuroki}}\ and\ \bibinfo {author} {\bibfnamefont {R.}~\bibnamefont {Arita}},\
  }\href {\doibase 10.1143/JPSJ.76.083707} {\bibfield  {journal} {\bibinfo
  {journal} {J. Phys. Soc. Jpn.}\ }\textbf {\bibinfo {volume} {76}},\ \bibinfo
  {pages} {083707} (\bibinfo {year} {2007})}\BibitemShut {NoStop}%
\bibitem [{\citenamefont {Hicks}\ and\ \citenamefont
  {Dresselhaus}(1993{\natexlab{a}})}]{hicks_effect_1993}%
  \BibitemOpen
  \bibfield  {author} {\bibinfo {author} {\bibfnamefont {L.~D.}\ \bibnamefont
  {Hicks}}\ and\ \bibinfo {author} {\bibfnamefont {M.~S.}\ \bibnamefont
  {Dresselhaus}},\ }\href {\doibase 10.1103/PhysRevB.47.12727} {\bibfield
  {journal} {\bibinfo  {journal} {Phys. Rev. B}\ }\textbf {\bibinfo {volume}
  {47}},\ \bibinfo {pages} {12727} (\bibinfo {year}
  {1993}{\natexlab{a}})}\BibitemShut {NoStop}%
\bibitem [{\citenamefont {Hicks}\ and\ \citenamefont
  {Dresselhaus}(1993{\natexlab{b}})}]{hicks_thermoelectric_1993}%
  \BibitemOpen
  \bibfield  {author} {\bibinfo {author} {\bibfnamefont {L.~D.}\ \bibnamefont
  {Hicks}}\ and\ \bibinfo {author} {\bibfnamefont {M.~S.}\ \bibnamefont
  {Dresselhaus}},\ }\href {\doibase 10.1103/PhysRevB.47.16631} {\bibfield
  {journal} {\bibinfo  {journal} {Phys. Rev. B}\ }\textbf {\bibinfo {volume}
  {47}},\ \bibinfo {pages} {16631} (\bibinfo {year}
  {1993}{\natexlab{b}})}\BibitemShut {NoStop}%
\bibitem [{\citenamefont {Usui}\ and\ \citenamefont
  {Kuroki}(2017)}]{usui_enhanced_2017}%
  \BibitemOpen
  \bibfield  {author} {\bibinfo {author} {\bibfnamefont {H.}~\bibnamefont
  {Usui}}\ and\ \bibinfo {author} {\bibfnamefont {K.}~\bibnamefont {Kuroki}},\
  }\href {\doibase 10.1063/1.4981890} {\bibfield  {journal} {\bibinfo
  {journal} {J. Appl. Phys.}\ }\textbf {\bibinfo {volume} {121}},\ \bibinfo
  {pages} {165101} (\bibinfo {year} {2017})}\BibitemShut {NoStop}%
\bibitem [{\citenamefont {Zhao}\ \emph {et~al.}(2014)\citenamefont {Zhao},
  \citenamefont {Lo}, \citenamefont {Zhang}, \citenamefont {Sun}, \citenamefont
  {Tan}, \citenamefont {Uher}, \citenamefont {Wolverton}, \citenamefont
  {Dravid},\ and\ \citenamefont {Kanatzidis}}]{zhao_ultralow_2014}%
  \BibitemOpen
  \bibfield  {author} {\bibinfo {author} {\bibfnamefont {L.-D.}\ \bibnamefont
  {Zhao}}, \bibinfo {author} {\bibfnamefont {S.-H.}\ \bibnamefont {Lo}},
  \bibinfo {author} {\bibfnamefont {Y.}~\bibnamefont {Zhang}}, \bibinfo
  {author} {\bibfnamefont {H.}~\bibnamefont {Sun}}, \bibinfo {author}
  {\bibfnamefont {G.}~\bibnamefont {Tan}}, \bibinfo {author} {\bibfnamefont
  {C.}~\bibnamefont {Uher}}, \bibinfo {author} {\bibfnamefont {C.}~\bibnamefont
  {Wolverton}}, \bibinfo {author} {\bibfnamefont {V.~P.}\ \bibnamefont
  {Dravid}}, \ and\ \bibinfo {author} {\bibfnamefont {M.~G.}\ \bibnamefont
  {Kanatzidis}},\ }\href {\doibase 10.1038/nature13184} {\bibfield  {journal}
  {\bibinfo  {journal} {Nature}\ }\textbf {\bibinfo {volume} {508}},\ \bibinfo
  {pages} {373} (\bibinfo {year} {2014})}\BibitemShut {NoStop}%
\bibitem [{\citenamefont {Kutorasinski}\ \emph {et~al.}(2015)\citenamefont
  {Kutorasinski}, \citenamefont {Wiendlocha}, \citenamefont {Kaprzyk},\ and\
  \citenamefont {Tobola}}]{kutorasinski_electronic_2015}%
  \BibitemOpen
  \bibfield  {author} {\bibinfo {author} {\bibfnamefont {K.}~\bibnamefont
  {Kutorasinski}}, \bibinfo {author} {\bibfnamefont {B.}~\bibnamefont
  {Wiendlocha}}, \bibinfo {author} {\bibfnamefont {S.}~\bibnamefont {Kaprzyk}},
  \ and\ \bibinfo {author} {\bibfnamefont {J.}~\bibnamefont {Tobola}},\ }\href
  {\doibase 10.1103/PhysRevB.91.205201} {\bibfield  {journal} {\bibinfo
  {journal} {Phys. Rev. B}\ }\textbf {\bibinfo {volume} {91}},\ \bibinfo
  {pages} {205201} (\bibinfo {year} {2015})}\BibitemShut {NoStop}%
\bibitem [{\citenamefont {Wang}\ \emph {et~al.}(2018)\citenamefont {Wang},
  \citenamefont {Fan}, \citenamefont {Shen}, \citenamefont {Hua}, \citenamefont
  {Hu}, \citenamefont {Sheng}, \citenamefont {Lu}, \citenamefont {Fang},
  \citenamefont {Qiu}, \citenamefont {Lu}, \citenamefont {Liu}, \citenamefont
  {Liu}, \citenamefont {Huang}, \citenamefont {Xu}, \citenamefont {Shen},\ and\
  \citenamefont {Zheng}}]{wang_defects_2018}%
  \BibitemOpen
  \bibfield  {author} {\bibinfo {author} {\bibfnamefont {Z.}~\bibnamefont
  {Wang}}, \bibinfo {author} {\bibfnamefont {C.}~\bibnamefont {Fan}}, \bibinfo
  {author} {\bibfnamefont {Z.}~\bibnamefont {Shen}}, \bibinfo {author}
  {\bibfnamefont {C.}~\bibnamefont {Hua}}, \bibinfo {author} {\bibfnamefont
  {Q.}~\bibnamefont {Hu}}, \bibinfo {author} {\bibfnamefont {F.}~\bibnamefont
  {Sheng}}, \bibinfo {author} {\bibfnamefont {Y.}~\bibnamefont {Lu}}, \bibinfo
  {author} {\bibfnamefont {H.}~\bibnamefont {Fang}}, \bibinfo {author}
  {\bibfnamefont {Z.}~\bibnamefont {Qiu}}, \bibinfo {author} {\bibfnamefont
  {J.}~\bibnamefont {Lu}}, \bibinfo {author} {\bibfnamefont {Z.}~\bibnamefont
  {Liu}}, \bibinfo {author} {\bibfnamefont {W.}~\bibnamefont {Liu}}, \bibinfo
  {author} {\bibfnamefont {Y.}~\bibnamefont {Huang}}, \bibinfo {author}
  {\bibfnamefont {Z.-A.}\ \bibnamefont {Xu}}, \bibinfo {author} {\bibfnamefont
  {D.~W.}\ \bibnamefont {Shen}}, \ and\ \bibinfo {author} {\bibfnamefont
  {Y.}~\bibnamefont {Zheng}},\ }\href {\doibase 10.1038/s41467-017-02566-1}
  {\bibfield  {journal} {\bibinfo  {journal} {Nat. Commun.}\ }\textbf {\bibinfo
  {volume} {9}},\ \bibinfo {pages} {47} (\bibinfo {year} {2018})}\BibitemShut
  {NoStop}%
\bibitem [{\citenamefont {Bilc}\ \emph {et~al.}(2015)\citenamefont {Bilc},
  \citenamefont {Hautier}, \citenamefont {Waroquiers}, \citenamefont
  {Rignanese},\ and\ \citenamefont {Ghosez}}]{bilc_low-dimensional_2015}%
  \BibitemOpen
  \bibfield  {author} {\bibinfo {author} {\bibfnamefont {D.~I.}\ \bibnamefont
  {Bilc}}, \bibinfo {author} {\bibfnamefont {G.}~\bibnamefont {Hautier}},
  \bibinfo {author} {\bibfnamefont {D.}~\bibnamefont {Waroquiers}}, \bibinfo
  {author} {\bibfnamefont {G.-M.}\ \bibnamefont {Rignanese}}, \ and\ \bibinfo
  {author} {\bibfnamefont {P.}~\bibnamefont {Ghosez}},\ }\href {\doibase
  10.1103/PhysRevLett.114.136601} {\bibfield  {journal} {\bibinfo  {journal}
  {Phys. Rev. Lett.}\ }\textbf {\bibinfo {volume} {114}},\ \bibinfo {pages}
  {136601} (\bibinfo {year} {2015})}\BibitemShut {NoStop}%
\bibitem [{\citenamefont {Lamontagne}\ \emph {et~al.}(2016)\citenamefont
  {Lamontagne}, \citenamefont {Laurita}, \citenamefont {Gaultois},
  \citenamefont {Knight}, \citenamefont {Ghadbeigi}, \citenamefont {Sparks},
  \citenamefont {Gruner}, \citenamefont {Pentcheva}, \citenamefont {Brown},\
  and\ \citenamefont {Seshadri}}]{lamontagne_high_2016}%
  \BibitemOpen
  \bibfield  {author} {\bibinfo {author} {\bibfnamefont {L.~K.}\ \bibnamefont
  {Lamontagne}}, \bibinfo {author} {\bibfnamefont {G.}~\bibnamefont {Laurita}},
  \bibinfo {author} {\bibfnamefont {M.~W.}\ \bibnamefont {Gaultois}}, \bibinfo
  {author} {\bibfnamefont {M.}~\bibnamefont {Knight}}, \bibinfo {author}
  {\bibfnamefont {L.}~\bibnamefont {Ghadbeigi}}, \bibinfo {author}
  {\bibfnamefont {T.~D.}\ \bibnamefont {Sparks}}, \bibinfo {author}
  {\bibfnamefont {M.~E.}\ \bibnamefont {Gruner}}, \bibinfo {author}
  {\bibfnamefont {R.}~\bibnamefont {Pentcheva}}, \bibinfo {author}
  {\bibfnamefont {C.~M.}\ \bibnamefont {Brown}}, \ and\ \bibinfo {author}
  {\bibfnamefont {R.}~\bibnamefont {Seshadri}},\ }\href {\doibase
  10.1021/acs.chemmater.6b00447} {\bibfield  {journal} {\bibinfo  {journal}
  {Chem. Mater.}\ }\textbf {\bibinfo {volume} {28}},\ \bibinfo {pages} {3367}
  (\bibinfo {year} {2016})}\BibitemShut {NoStop}%
\bibitem [{\citenamefont {He}\ \emph {et~al.}(2017)\citenamefont {He},
  \citenamefont {Hao}, \citenamefont {Xia}, \citenamefont {Naghavi},
  \citenamefont {Ozoli\c{n}\v{s}},\ and\ \citenamefont
  {Wolverton}}]{he_bi2pdo4_2017}%
  \BibitemOpen
  \bibfield  {author} {\bibinfo {author} {\bibfnamefont {J.}~\bibnamefont
  {He}}, \bibinfo {author} {\bibfnamefont {S.}~\bibnamefont {Hao}}, \bibinfo
  {author} {\bibfnamefont {Y.}~\bibnamefont {Xia}}, \bibinfo {author}
  {\bibfnamefont {S.~S.}\ \bibnamefont {Naghavi}}, \bibinfo {author}
  {\bibfnamefont {V.}~\bibnamefont {Ozoli\c{n}\v{s}}}, \ and\ \bibinfo {author}
  {\bibfnamefont {C.}~\bibnamefont {Wolverton}},\ }\href {\doibase
  10.1021/acs.chemmater.6b04230} {\bibfield  {journal} {\bibinfo  {journal}
  {Chem. Mater.}\ }\textbf {\bibinfo {volume} {29}},\ \bibinfo {pages} {2529}
  (\bibinfo {year} {2017})}\BibitemShut {NoStop}%
\bibitem [{\citenamefont {Isaacs}\ and\ \citenamefont
  {Wolverton}(2018)}]{isaacs_inverse_2018}%
  \BibitemOpen
  \bibfield  {author} {\bibinfo {author} {\bibfnamefont {E.~B.}\ \bibnamefont
  {Isaacs}}\ and\ \bibinfo {author} {\bibfnamefont {C.}~\bibnamefont
  {Wolverton}},\ }\href {\doibase 10.1021/acs.chemmater.7b04496} {\bibfield
  {journal} {\bibinfo  {journal} {Chem. Mater.}\ }\textbf {\bibinfo {volume}
  {30}},\ \bibinfo {pages} {1540} (\bibinfo {year} {2018})}\BibitemShut
  {NoStop}%
\bibitem [{\citenamefont {He}\ \emph {et~al.}(2011)\citenamefont {He},
  \citenamefont {Liu},\ and\ \citenamefont {Funahashi}}]{he_oxide_2011}%
  \BibitemOpen
  \bibfield  {author} {\bibinfo {author} {\bibfnamefont {J.}~\bibnamefont
  {He}}, \bibinfo {author} {\bibfnamefont {Y.}~\bibnamefont {Liu}}, \ and\
  \bibinfo {author} {\bibfnamefont {R.}~\bibnamefont {Funahashi}},\ }\href
  {\doibase 10.1557/jmr.2011.108} {\bibfield  {journal} {\bibinfo  {journal}
  {J. Mater. Res.}\ }\textbf {\bibinfo {volume} {26}},\ \bibinfo {pages} {1762}
  (\bibinfo {year} {2011})}\BibitemShut {NoStop}%
\bibitem [{\citenamefont {Garrity}(2016)}]{garrity_first-principles_2016}%
  \BibitemOpen
  \bibfield  {author} {\bibinfo {author} {\bibfnamefont {K.~F.}\ \bibnamefont
  {Garrity}},\ }\href {\doibase 10.1103/PhysRevB.94.045122} {\bibfield
  {journal} {\bibinfo  {journal} {Phys. Rev. B}\ }\textbf {\bibinfo {volume}
  {94}},\ \bibinfo {pages} {045122} (\bibinfo {year} {2016})}\BibitemShut
  {NoStop}%
\bibitem [{\citenamefont {Snyder}\ and\ \citenamefont
  {Toberer}(2008)}]{snyder_complex_2008}%
  \BibitemOpen
  \bibfield  {author} {\bibinfo {author} {\bibfnamefont {G.~J.}\ \bibnamefont
  {Snyder}}\ and\ \bibinfo {author} {\bibfnamefont {E.~S.}\ \bibnamefont
  {Toberer}},\ }\href {\doibase 10.1038/nmat2090} {\bibfield  {journal}
  {\bibinfo  {journal} {Nat. Mater.}\ }\textbf {\bibinfo {volume} {7}},\
  \bibinfo {pages} {105} (\bibinfo {year} {2008})}\BibitemShut {NoStop}%
\bibitem [{\citenamefont {Zhang}\ and\ \citenamefont
  {Zhao}(2015)}]{zhang_thermoelectric_2015}%
  \BibitemOpen
  \bibfield  {author} {\bibinfo {author} {\bibfnamefont {X.}~\bibnamefont
  {Zhang}}\ and\ \bibinfo {author} {\bibfnamefont {L.-D.}\ \bibnamefont
  {Zhao}},\ }\href {\doibase 10.1016/j.jmat.2015.01.001} {\bibfield  {journal}
  {\bibinfo  {journal} {J. Materiomics}\ }\textbf {\bibinfo {volume} {1}},\
  \bibinfo {pages} {92} (\bibinfo {year} {2015})}\BibitemShut {NoStop}%
\bibitem [{\citenamefont {Tan}\ \emph {et~al.}(2016)\citenamefont {Tan},
  \citenamefont {Zhao},\ and\ \citenamefont
  {Kanatzidis}}]{tan_rationally_2016}%
  \BibitemOpen
  \bibfield  {author} {\bibinfo {author} {\bibfnamefont {G.}~\bibnamefont
  {Tan}}, \bibinfo {author} {\bibfnamefont {L.-D.}\ \bibnamefont {Zhao}}, \
  and\ \bibinfo {author} {\bibfnamefont {M.~G.}\ \bibnamefont {Kanatzidis}},\
  }\href {\doibase 10.1021/acs.chemrev.6b00255} {\bibfield  {journal} {\bibinfo
   {journal} {Chem. Rev.}\ }\textbf {\bibinfo {volume} {116}},\ \bibinfo
  {pages} {12123} (\bibinfo {year} {2016})}\BibitemShut {NoStop}%
\bibitem [{\citenamefont {Yao}\ \emph {et~al.}(2004)\citenamefont {Yao},
  \citenamefont {Wang},\ and\ \citenamefont {Liu}}]{yao_first_2004}%
  \BibitemOpen
  \bibfield  {author} {\bibinfo {author} {\bibfnamefont {K.~L.}\ \bibnamefont
  {Yao}}, \bibinfo {author} {\bibfnamefont {L.~Q.}\ \bibnamefont {Wang}}, \
  and\ \bibinfo {author} {\bibfnamefont {Z.~L.}\ \bibnamefont {Liu}},\ }\href
  {\doibase 10.1016/j.ssc.2004.03.053} {\bibfield  {journal} {\bibinfo
  {journal} {Solid State Comm.}\ }\textbf {\bibinfo {volume} {130}},\ \bibinfo
  {pages} {741} (\bibinfo {year} {2004})}\BibitemShut {NoStop}%
\bibitem [{\citenamefont {Bhatt}\ \emph {et~al.}(2013)\citenamefont {Bhatt},
  \citenamefont {Bhattacharya}, \citenamefont {Basu}, \citenamefont {Singh},
  \citenamefont {Deshpande}, \citenamefont {Surger}, \citenamefont {Basu},
  \citenamefont {Aswal},\ and\ \citenamefont {Gupta}}]{bhatt_growth_2013}%
  \BibitemOpen
  \bibfield  {author} {\bibinfo {author} {\bibfnamefont {R.}~\bibnamefont
  {Bhatt}}, \bibinfo {author} {\bibfnamefont {S.}~\bibnamefont {Bhattacharya}},
  \bibinfo {author} {\bibfnamefont {R.}~\bibnamefont {Basu}}, \bibinfo {author}
  {\bibfnamefont {A.}~\bibnamefont {Singh}}, \bibinfo {author} {\bibfnamefont
  {U.}~\bibnamefont {Deshpande}}, \bibinfo {author} {\bibfnamefont
  {C.}~\bibnamefont {Surger}}, \bibinfo {author} {\bibfnamefont
  {S.}~\bibnamefont {Basu}}, \bibinfo {author} {\bibfnamefont {D.~K.}\
  \bibnamefont {Aswal}}, \ and\ \bibinfo {author} {\bibfnamefont {S.~K.}\
  \bibnamefont {Gupta}},\ }\href {\doibase 10.1016/j.tsf.2013.04.143}
  {\bibfield  {journal} {\bibinfo  {journal} {Thin Solid Films}\ }\textbf
  {\bibinfo {volume} {539}},\ \bibinfo {pages} {41} (\bibinfo {year}
  {2013})}\BibitemShut {NoStop}%
\bibitem [{\citenamefont {Sun}\ \emph {et~al.}(2015)\citenamefont {Sun},
  \citenamefont {Shi}, \citenamefont {Siegrist},\ and\ \citenamefont
  {Singh}}]{sun_electronic_2015}%
  \BibitemOpen
  \bibfield  {author} {\bibinfo {author} {\bibfnamefont {J.}~\bibnamefont
  {Sun}}, \bibinfo {author} {\bibfnamefont {H.}~\bibnamefont {Shi}}, \bibinfo
  {author} {\bibfnamefont {T.}~\bibnamefont {Siegrist}}, \ and\ \bibinfo
  {author} {\bibfnamefont {D.~J.}\ \bibnamefont {Singh}},\ }\href {\doibase
  10.1063/1.4933302} {\bibfield  {journal} {\bibinfo  {journal} {Appl. Phys.
  Lett.}\ }\textbf {\bibinfo {volume} {107}},\ \bibinfo {pages} {153902}
  (\bibinfo {year} {2015})}\BibitemShut {NoStop}%
\bibitem [{\citenamefont {Huster}\ and\ \citenamefont
  {Bronger}(1986)}]{huster_synthesis_1986}%
  \BibitemOpen
  \bibfield  {author} {\bibinfo {author} {\bibfnamefont {J.}~\bibnamefont
  {Huster}}\ and\ \bibinfo {author} {\bibfnamefont {W.}~\bibnamefont
  {Bronger}},\ }\href {\doibase 10.1016/0022-5088(86)90206-7} {\bibfield
  {journal} {\bibinfo  {journal} {J. Less Common Met.}\ }\textbf {\bibinfo
  {volume} {119}},\ \bibinfo {pages} {159} (\bibinfo {year}
  {1986})}\BibitemShut {NoStop}%
\bibitem [{\citenamefont {Madsen}(2006)}]{madsen_automated_2006}%
  \BibitemOpen
  \bibfield  {author} {\bibinfo {author} {\bibfnamefont {G.~K.~H.}\
  \bibnamefont {Madsen}},\ }\href {\doibase 10.1021/ja062526a} {\bibfield
  {journal} {\bibinfo  {journal} {J. Am. Chem. Soc.}\ }\textbf {\bibinfo
  {volume} {128}},\ \bibinfo {pages} {12140} (\bibinfo {year}
  {2006})}\BibitemShut {NoStop}%
\bibitem [{\citenamefont {Gandi}\ and\ \citenamefont
  {Schwingenschlögl}(2014)}]{gandi_ws2_2014}%
  \BibitemOpen
  \bibfield  {author} {\bibinfo {author} {\bibfnamefont {A.~N.}\ \bibnamefont
  {Gandi}}\ and\ \bibinfo {author} {\bibfnamefont {U.}~\bibnamefont
  {Schwingenschlögl}},\ }\href {\doibase 10.1021/cm503487n} {\bibfield
  {journal} {\bibinfo  {journal} {Chem. Mater.}\ }\textbf {\bibinfo {volume}
  {26}},\ \bibinfo {pages} {6628} (\bibinfo {year} {2014})}\BibitemShut
  {NoStop}%
\bibitem [{\citenamefont {Wang}\ \emph {et~al.}(2015)\citenamefont {Wang},
  \citenamefont {Zhang}, \citenamefont {Yu},\ and\ \citenamefont
  {Wang}}]{wang_thermoelectric_2015}%
  \BibitemOpen
  \bibfield  {author} {\bibinfo {author} {\bibfnamefont {F.~Q.}\ \bibnamefont
  {Wang}}, \bibinfo {author} {\bibfnamefont {S.}~\bibnamefont {Zhang}},
  \bibinfo {author} {\bibfnamefont {J.}~\bibnamefont {Yu}}, \ and\ \bibinfo
  {author} {\bibfnamefont {Q.}~\bibnamefont {Wang}},\ }\href {\doibase
  10.1039/C5NR03813H} {\bibfield  {journal} {\bibinfo  {journal} {Nanoscale}\
  }\textbf {\bibinfo {volume} {7}},\ \bibinfo {pages} {15962} (\bibinfo {year}
  {2015})}\BibitemShut {NoStop}%
\bibitem [{\citenamefont {Hao}\ \emph {et~al.}(2016)\citenamefont {Hao},
  \citenamefont {Shi}, \citenamefont {Dravid}, \citenamefont {Kanatzidis},\
  and\ \citenamefont {Wolverton}}]{hao_computational_2016}%
  \BibitemOpen
  \bibfield  {author} {\bibinfo {author} {\bibfnamefont {S.}~\bibnamefont
  {Hao}}, \bibinfo {author} {\bibfnamefont {F.}~\bibnamefont {Shi}}, \bibinfo
  {author} {\bibfnamefont {V.~P.}\ \bibnamefont {Dravid}}, \bibinfo {author}
  {\bibfnamefont {M.~G.}\ \bibnamefont {Kanatzidis}}, \ and\ \bibinfo {author}
  {\bibfnamefont {C.}~\bibnamefont {Wolverton}},\ }\href {\doibase
  10.1021/acs.chemmater.6b01164} {\bibfield  {journal} {\bibinfo  {journal}
  {Chem. Mater.}\ }\textbf {\bibinfo {volume} {28}},\ \bibinfo {pages} {3218}
  (\bibinfo {year} {2016})}\BibitemShut {NoStop}%
\bibitem [{\citenamefont {Pal}\ \emph {et~al.}(2018)\citenamefont {Pal},
  \citenamefont {He},\ and\ \citenamefont {Wolverton}}]{pal_bonding_2018}%
  \BibitemOpen
  \bibfield  {author} {\bibinfo {author} {\bibfnamefont {K.}~\bibnamefont
  {Pal}}, \bibinfo {author} {\bibfnamefont {J.}~\bibnamefont {He}}, \ and\
  \bibinfo {author} {\bibfnamefont {C.}~\bibnamefont {Wolverton}},\ }\href
  {\doibase 10.1021/acs.chemmater.8b03130} {\bibfield  {journal} {\bibinfo
  {journal} {Chem. Mater.}\ } (\bibinfo {year} {2018}),\
  10.1021/acs.chemmater.8b03130}\BibitemShut {NoStop}%
\bibitem [{\citenamefont {Hohenberg}\ and\ \citenamefont
  {Kohn}(1964)}]{hohenberg_inhomogeneous_1964}%
  \BibitemOpen
  \bibfield  {author} {\bibinfo {author} {\bibfnamefont {P.}~\bibnamefont
  {Hohenberg}}\ and\ \bibinfo {author} {\bibfnamefont {W.}~\bibnamefont
  {Kohn}},\ }\href@noop {} {\bibfield  {journal} {\bibinfo  {journal} {Phys.
  Rev.}\ }\textbf {\bibinfo {volume} {136}},\ \bibinfo {pages} {B864} (\bibinfo
  {year} {1964})}\BibitemShut {NoStop}%
\bibitem [{\citenamefont {Kohn}\ and\ \citenamefont
  {Sham}(1965)}]{kohn_self-consistent_1965}%
  \BibitemOpen
  \bibfield  {author} {\bibinfo {author} {\bibfnamefont {W.}~\bibnamefont
  {Kohn}}\ and\ \bibinfo {author} {\bibfnamefont {L.~J.}\ \bibnamefont
  {Sham}},\ }\href@noop {} {\bibfield  {journal} {\bibinfo  {journal} {Phys.
  Rev.}\ }\textbf {\bibinfo {volume} {140}},\ \bibinfo {pages} {A1133}
  (\bibinfo {year} {1965})}\BibitemShut {NoStop}%
\bibitem [{\citenamefont {Perdew}\ \emph {et~al.}(1996)\citenamefont {Perdew},
  \citenamefont {Burke},\ and\ \citenamefont
  {Ernzerhof}}]{perdew_generalized_1996}%
  \BibitemOpen
  \bibfield  {author} {\bibinfo {author} {\bibfnamefont {J.~P.}\ \bibnamefont
  {Perdew}}, \bibinfo {author} {\bibfnamefont {K.}~\bibnamefont {Burke}}, \
  and\ \bibinfo {author} {\bibfnamefont {M.}~\bibnamefont {Ernzerhof}},\
  }\href@noop {} {\bibfield  {journal} {\bibinfo  {journal} {Phys. Rev. Lett.}\
  }\textbf {\bibinfo {volume} {77}},\ \bibinfo {pages} {3865} (\bibinfo {year}
  {1996})}\BibitemShut {NoStop}%
\bibitem [{\citenamefont {Bl{\"o}chl}(1994)}]{blochl_projector_1994}%
  \BibitemOpen
  \bibfield  {author} {\bibinfo {author} {\bibfnamefont {P.~E.}\ \bibnamefont
  {Bl{\"o}chl}},\ }\href@noop {} {\bibfield  {journal} {\bibinfo  {journal}
  {Phys. Rev. B}\ }\textbf {\bibinfo {volume} {50}},\ \bibinfo {pages} {17953}
  (\bibinfo {year} {1994})}\BibitemShut {NoStop}%
\bibitem [{\citenamefont {Kresse}\ and\ \citenamefont
  {Joubert}(1999)}]{kresse_ultrasoft_1999}%
  \BibitemOpen
  \bibfield  {author} {\bibinfo {author} {\bibfnamefont {G.}~\bibnamefont
  {Kresse}}\ and\ \bibinfo {author} {\bibfnamefont {D.}~\bibnamefont
  {Joubert}},\ }\href@noop {} {\bibfield  {journal} {\bibinfo  {journal} {Phys.
  Rev. B}\ }\textbf {\bibinfo {volume} {59}},\ \bibinfo {pages} {1758}
  (\bibinfo {year} {1999})}\BibitemShut {NoStop}%
\bibitem [{\citenamefont {Kresse}\ and\ \citenamefont
  {Hafner}(1994)}]{kresse_ab_1994}%
  \BibitemOpen
  \bibfield  {author} {\bibinfo {author} {\bibfnamefont {G.}~\bibnamefont
  {Kresse}}\ and\ \bibinfo {author} {\bibfnamefont {J.}~\bibnamefont
  {Hafner}},\ }\href@noop {} {\bibfield  {journal} {\bibinfo  {journal} {Phys.
  Rev. B}\ }\textbf {\bibinfo {volume} {49}},\ \bibinfo {pages} {14251}
  (\bibinfo {year} {1994})}\BibitemShut {NoStop}%
\bibitem [{\citenamefont {Kresse}\ and\ \citenamefont
  {Hafner}(1993)}]{kresse_ab_1993}%
  \BibitemOpen
  \bibfield  {author} {\bibinfo {author} {\bibfnamefont {G.}~\bibnamefont
  {Kresse}}\ and\ \bibinfo {author} {\bibfnamefont {J.}~\bibnamefont
  {Hafner}},\ }\href@noop {} {\bibfield  {journal} {\bibinfo  {journal} {Phys.
  Rev. B}\ }\textbf {\bibinfo {volume} {47}},\ \bibinfo {pages} {558} (\bibinfo
  {year} {1993})}\BibitemShut {NoStop}%
\bibitem [{\citenamefont {Kresse}\ and\ \citenamefont
  {Furthm{\"u}ller}(1996{\natexlab{a}})}]{kresse_efficient_1996}%
  \BibitemOpen
  \bibfield  {author} {\bibinfo {author} {\bibfnamefont {G.}~\bibnamefont
  {Kresse}}\ and\ \bibinfo {author} {\bibfnamefont {J.}~\bibnamefont
  {Furthm{\"u}ller}},\ }\href@noop {} {\bibfield  {journal} {\bibinfo
  {journal} {Phys. Rev. B}\ }\textbf {\bibinfo {volume} {54}},\ \bibinfo
  {pages} {11169} (\bibinfo {year} {1996}{\natexlab{a}})}\BibitemShut {NoStop}%
\bibitem [{\citenamefont {Kresse}\ and\ \citenamefont
  {Furthm{\"u}ller}(1996{\natexlab{b}})}]{kresse_efficiency_1996}%
  \BibitemOpen
  \bibfield  {author} {\bibinfo {author} {\bibfnamefont {G.}~\bibnamefont
  {Kresse}}\ and\ \bibinfo {author} {\bibfnamefont {J.}~\bibnamefont
  {Furthm{\"u}ller}},\ }\href@noop {} {\bibfield  {journal} {\bibinfo
  {journal} {Comput. Mater. Sci.}\ }\textbf {\bibinfo {volume} {6}},\ \bibinfo
  {pages} {15} (\bibinfo {year} {1996}{\natexlab{b}})}\BibitemShut {NoStop}%
\bibitem [{\citenamefont {Setyawan}\ and\ \citenamefont
  {Curtarolo}(2010)}]{setyawan_high-throughput_2010}%
  \BibitemOpen
  \bibfield  {author} {\bibinfo {author} {\bibfnamefont {W.}~\bibnamefont
  {Setyawan}}\ and\ \bibinfo {author} {\bibfnamefont {S.}~\bibnamefont
  {Curtarolo}},\ }\href {\doibase 10.1016/j.commatsci.2010.05.010} {\bibfield
  {journal} {\bibinfo  {journal} {Comput. Mater. Sci.}\ }\textbf {\bibinfo
  {volume} {49}},\ \bibinfo {pages} {299} (\bibinfo {year} {2010})}\BibitemShut
  {NoStop}%
\bibitem [{\citenamefont {Madsen}\ and\ \citenamefont
  {Singh}(2006)}]{madsen_boltztrap_2006}%
  \BibitemOpen
  \bibfield  {author} {\bibinfo {author} {\bibfnamefont {G.~K.~H.}\
  \bibnamefont {Madsen}}\ and\ \bibinfo {author} {\bibfnamefont {D.~J.}\
  \bibnamefont {Singh}},\ }\href {\doibase 10.1016/j.cpc.2006.03.007}
  {\bibfield  {journal} {\bibinfo  {journal} {Comput. Phys. Commun.}\ }\textbf
  {\bibinfo {volume} {175}},\ \bibinfo {pages} {67} (\bibinfo {year}
  {2006})}\BibitemShut {NoStop}%
\bibitem [{\citenamefont {Togo}\ \emph {et~al.}(2008)\citenamefont {Togo},
  \citenamefont {Oba},\ and\ \citenamefont {Tanaka}}]{phonopy}%
  \BibitemOpen
  \bibfield  {author} {\bibinfo {author} {\bibfnamefont {A.}~\bibnamefont
  {Togo}}, \bibinfo {author} {\bibfnamefont {F.}~\bibnamefont {Oba}}, \ and\
  \bibinfo {author} {\bibfnamefont {I.}~\bibnamefont {Tanaka}},\ }\href@noop {}
  {\bibfield  {journal} {\bibinfo  {journal} {Phys. Rev. B}\ }\textbf {\bibinfo
  {volume} {78}},\ \bibinfo {pages} {134106} (\bibinfo {year}
  {2008})}\BibitemShut {NoStop}%
\bibitem [{\citenamefont {Erhart}\ \emph {et~al.}(2015)\citenamefont {Erhart},
  \citenamefont {Sadigh}, \citenamefont {Schleife},\ and\ \citenamefont
  {\r{A}berg}}]{erhart_first-principles_2015}%
  \BibitemOpen
  \bibfield  {author} {\bibinfo {author} {\bibfnamefont {P.}~\bibnamefont
  {Erhart}}, \bibinfo {author} {\bibfnamefont {B.}~\bibnamefont {Sadigh}},
  \bibinfo {author} {\bibfnamefont {A.}~\bibnamefont {Schleife}}, \ and\
  \bibinfo {author} {\bibfnamefont {D.}~\bibnamefont {\r{A}berg}},\ }\href
  {\doibase 10.1103/PhysRevB.91.165206} {\bibfield  {journal} {\bibinfo
  {journal} {Phys. Rev. B}\ }\textbf {\bibinfo {volume} {91}},\ \bibinfo
  {pages} {165206} (\bibinfo {year} {2015})}\BibitemShut {NoStop}%
\bibitem [{\citenamefont {Larsen}\ \emph {et~al.}(2017)\citenamefont {Larsen},
  \citenamefont {Mortensen}, \citenamefont {Blomqvist}, \citenamefont
  {Castelli}, \citenamefont {Christensen}, \citenamefont {Du{\l}ak},
  \citenamefont {Friis}, \citenamefont {Groves}, \citenamefont {Hammer},
  \citenamefont {Hargus}, \citenamefont {Hermes}, \citenamefont {Jennings},
  \citenamefont {Jensen}, \citenamefont {Kermode}, \citenamefont {Kitchin},
  \citenamefont {Kolsbjerg}, \citenamefont {Kubal}, \citenamefont {Kaasbjerg},
  \citenamefont {Lysgaard}, \citenamefont {Maronsson}, \citenamefont {Maxson},
  \citenamefont {Olsen}, \citenamefont {Pastewka}, \citenamefont {Peterson},
  \citenamefont {Rostgaard}, \citenamefont {Schi{\o}tz}, \citenamefont
  {Sch\"{u}tt}, \citenamefont {Strange}, \citenamefont {Thygesen},
  \citenamefont {Vegge}, \citenamefont {Vilhelmsen}, \citenamefont {Walter},
  \citenamefont {Zeng},\ and\ \citenamefont {Jacobsen}}]{ase-paper}%
  \BibitemOpen
  \bibfield  {author} {\bibinfo {author} {\bibfnamefont {A.~H.}\ \bibnamefont
  {Larsen}}, \bibinfo {author} {\bibfnamefont {J.~J.}\ \bibnamefont
  {Mortensen}}, \bibinfo {author} {\bibfnamefont {J.}~\bibnamefont
  {Blomqvist}}, \bibinfo {author} {\bibfnamefont {I.~E.}\ \bibnamefont
  {Castelli}}, \bibinfo {author} {\bibfnamefont {R.}~\bibnamefont
  {Christensen}}, \bibinfo {author} {\bibfnamefont {M.}~\bibnamefont
  {Du{\l}ak}}, \bibinfo {author} {\bibfnamefont {J.}~\bibnamefont {Friis}},
  \bibinfo {author} {\bibfnamefont {M.~N.}\ \bibnamefont {Groves}}, \bibinfo
  {author} {\bibfnamefont {B.}~\bibnamefont {Hammer}}, \bibinfo {author}
  {\bibfnamefont {C.}~\bibnamefont {Hargus}}, \bibinfo {author} {\bibfnamefont
  {E.~D.}\ \bibnamefont {Hermes}}, \bibinfo {author} {\bibfnamefont {P.~C.}\
  \bibnamefont {Jennings}}, \bibinfo {author} {\bibfnamefont {P.~B.}\
  \bibnamefont {Jensen}}, \bibinfo {author} {\bibfnamefont {J.}~\bibnamefont
  {Kermode}}, \bibinfo {author} {\bibfnamefont {J.~R.}\ \bibnamefont
  {Kitchin}}, \bibinfo {author} {\bibfnamefont {E.~L.}\ \bibnamefont
  {Kolsbjerg}}, \bibinfo {author} {\bibfnamefont {J.}~\bibnamefont {Kubal}},
  \bibinfo {author} {\bibfnamefont {K.}~\bibnamefont {Kaasbjerg}}, \bibinfo
  {author} {\bibfnamefont {S.}~\bibnamefont {Lysgaard}}, \bibinfo {author}
  {\bibfnamefont {J.~B.}\ \bibnamefont {Maronsson}}, \bibinfo {author}
  {\bibfnamefont {T.}~\bibnamefont {Maxson}}, \bibinfo {author} {\bibfnamefont
  {T.}~\bibnamefont {Olsen}}, \bibinfo {author} {\bibfnamefont
  {L.}~\bibnamefont {Pastewka}}, \bibinfo {author} {\bibfnamefont
  {A.}~\bibnamefont {Peterson}}, \bibinfo {author} {\bibfnamefont
  {C.}~\bibnamefont {Rostgaard}}, \bibinfo {author} {\bibfnamefont
  {J.}~\bibnamefont {Schi{\o}tz}}, \bibinfo {author} {\bibfnamefont
  {O.}~\bibnamefont {Sch\"{u}tt}}, \bibinfo {author} {\bibfnamefont
  {M.}~\bibnamefont {Strange}}, \bibinfo {author} {\bibfnamefont {K.~S.}\
  \bibnamefont {Thygesen}}, \bibinfo {author} {\bibfnamefont {T.}~\bibnamefont
  {Vegge}}, \bibinfo {author} {\bibfnamefont {L.}~\bibnamefont {Vilhelmsen}},
  \bibinfo {author} {\bibfnamefont {M.}~\bibnamefont {Walter}}, \bibinfo
  {author} {\bibfnamefont {Z.}~\bibnamefont {Zeng}}, \ and\ \bibinfo {author}
  {\bibfnamefont {K.~W.}\ \bibnamefont {Jacobsen}},\ }\href
  {http://stacks.iop.org/0953-8984/29/i=27/a=273002} {\bibfield  {journal}
  {\bibinfo  {journal} {J. Phys. Condens. Matter}\ }\textbf {\bibinfo {volume}
  {29}},\ \bibinfo {pages} {273002} (\bibinfo {year} {2017})}\BibitemShut
  {NoStop}%
\bibitem [{\citenamefont {Callaway}(1959)}]{callaway_model_1959}%
  \BibitemOpen
  \bibfield  {author} {\bibinfo {author} {\bibfnamefont {J.}~\bibnamefont
  {Callaway}},\ }\href {\doibase 10.1103/PhysRev.113.1046} {\bibfield
  {journal} {\bibinfo  {journal} {Phys. Rev.}\ }\textbf {\bibinfo {volume}
  {113}},\ \bibinfo {pages} {1046} (\bibinfo {year} {1959})}\BibitemShut
  {NoStop}%
\bibitem [{\citenamefont {Morelli}\ \emph {et~al.}(2002)\citenamefont
  {Morelli}, \citenamefont {Heremans},\ and\ \citenamefont
  {Slack}}]{morelli_estimation_2002}%
  \BibitemOpen
  \bibfield  {author} {\bibinfo {author} {\bibfnamefont {D.~T.}\ \bibnamefont
  {Morelli}}, \bibinfo {author} {\bibfnamefont {J.~P.}\ \bibnamefont
  {Heremans}}, \ and\ \bibinfo {author} {\bibfnamefont {G.~A.}\ \bibnamefont
  {Slack}},\ }\href {\doibase 10.1103/PhysRevB.66.195304} {\bibfield  {journal}
  {\bibinfo  {journal} {Phys. Rev. B}\ }\textbf {\bibinfo {volume} {66}},\
  \bibinfo {pages} {195304} (\bibinfo {year} {2002})}\BibitemShut {NoStop}%
\bibitem [{\citenamefont {Zhang}(2016)}]{zhang_first-principles_2016}%
  \BibitemOpen
  \bibfield  {author} {\bibinfo {author} {\bibfnamefont {Y.}~\bibnamefont
  {Zhang}},\ }\href {\doibase 10.1016/j.jmat.2016.06.004} {\bibfield  {journal}
  {\bibinfo  {journal} {J. Materiomics}\ }\textbf {\bibinfo {volume} {2}},\
  \bibinfo {pages} {237} (\bibinfo {year} {2016})}\BibitemShut {NoStop}%
\bibitem [{\citenamefont {Cahill}\ \emph {et~al.}(1992)\citenamefont {Cahill},
  \citenamefont {Watson},\ and\ \citenamefont {Pohl}}]{cahill_lower_1992}%
  \BibitemOpen
  \bibfield  {author} {\bibinfo {author} {\bibfnamefont {D.~G.}\ \bibnamefont
  {Cahill}}, \bibinfo {author} {\bibfnamefont {S.~K.}\ \bibnamefont {Watson}},
  \ and\ \bibinfo {author} {\bibfnamefont {R.~O.}\ \bibnamefont {Pohl}},\
  }\href {\doibase 10.1103/PhysRevB.46.6131} {\bibfield  {journal} {\bibinfo
  {journal} {Phys. Rev. B}\ }\textbf {\bibinfo {volume} {46}},\ \bibinfo
  {pages} {6131} (\bibinfo {year} {1992})}\BibitemShut {NoStop}%
\bibitem [{Note1()}]{Note1}%
  \BibitemOpen
  \bibinfo {note} {See Supplemental Material for additional details on
  electronic and thermal transport methods, phonon and anharmonicity
  calculation convergence, animations of several low-frequency phonon modes,
  comparison to experimental crystal structure and formation energy, and the
  electronic band structure for a different high-symmetry crystal momentum
  path.}\BibitemShut {Stop}%
\bibitem [{\citenamefont {Zhao}\ \emph {et~al.}(2015)\citenamefont {Zhao},
  \citenamefont {Tan}, \citenamefont {Hao}, \citenamefont {He}, \citenamefont
  {Pei}, \citenamefont {Chi}, \citenamefont {Wang}, \citenamefont {Gong},
  \citenamefont {Xu}, \citenamefont {Dravid}, \citenamefont {Uher},
  \citenamefont {Snyder}, \citenamefont {Wolverton},\ and\ \citenamefont
  {Kanatzidis}}]{zhao_ultrahigh_2015}%
  \BibitemOpen
  \bibfield  {author} {\bibinfo {author} {\bibfnamefont {L.-D.}\ \bibnamefont
  {Zhao}}, \bibinfo {author} {\bibfnamefont {G.}~\bibnamefont {Tan}}, \bibinfo
  {author} {\bibfnamefont {S.}~\bibnamefont {Hao}}, \bibinfo {author}
  {\bibfnamefont {J.}~\bibnamefont {He}}, \bibinfo {author} {\bibfnamefont
  {Y.}~\bibnamefont {Pei}}, \bibinfo {author} {\bibfnamefont {H.}~\bibnamefont
  {Chi}}, \bibinfo {author} {\bibfnamefont {H.}~\bibnamefont {Wang}}, \bibinfo
  {author} {\bibfnamefont {S.}~\bibnamefont {Gong}}, \bibinfo {author}
  {\bibfnamefont {H.}~\bibnamefont {Xu}}, \bibinfo {author} {\bibfnamefont
  {V.~P.}\ \bibnamefont {Dravid}}, \bibinfo {author} {\bibfnamefont
  {C.}~\bibnamefont {Uher}}, \bibinfo {author} {\bibfnamefont {G.~J.}\
  \bibnamefont {Snyder}}, \bibinfo {author} {\bibfnamefont {C.}~\bibnamefont
  {Wolverton}}, \ and\ \bibinfo {author} {\bibfnamefont {M.~G.}\ \bibnamefont
  {Kanatzidis}},\ }\href {\doibase 10.1126/science.aad3749} {\bibfield
  {journal} {\bibinfo  {journal} {Science}\ }\textbf {\bibinfo {volume}
  {351}},\ \bibinfo {pages} {141} (\bibinfo {year} {2015})}\BibitemShut
  {NoStop}%
\bibitem [{\citenamefont {Duong}\ \emph {et~al.}(2016)\citenamefont {Duong},
  \citenamefont {Nguyen}, \citenamefont {Duvjir}, \citenamefont {Duong},
  \citenamefont {Kwon}, \citenamefont {Song}, \citenamefont {Lee},
  \citenamefont {Lee}, \citenamefont {Park}, \citenamefont {Min}, \citenamefont
  {Lee}, \citenamefont {Kim},\ and\ \citenamefont
  {Cho}}]{duong_achieving_2016}%
  \BibitemOpen
  \bibfield  {author} {\bibinfo {author} {\bibfnamefont {A.~T.}\ \bibnamefont
  {Duong}}, \bibinfo {author} {\bibfnamefont {V.~Q.}\ \bibnamefont {Nguyen}},
  \bibinfo {author} {\bibfnamefont {G.}~\bibnamefont {Duvjir}}, \bibinfo
  {author} {\bibfnamefont {V.~T.}\ \bibnamefont {Duong}}, \bibinfo {author}
  {\bibfnamefont {S.}~\bibnamefont {Kwon}}, \bibinfo {author} {\bibfnamefont
  {J.~Y.}\ \bibnamefont {Song}}, \bibinfo {author} {\bibfnamefont {J.~K.}\
  \bibnamefont {Lee}}, \bibinfo {author} {\bibfnamefont {J.~E.}\ \bibnamefont
  {Lee}}, \bibinfo {author} {\bibfnamefont {S.}~\bibnamefont {Park}}, \bibinfo
  {author} {\bibfnamefont {T.}~\bibnamefont {Min}}, \bibinfo {author}
  {\bibfnamefont {J.}~\bibnamefont {Lee}}, \bibinfo {author} {\bibfnamefont
  {J.}~\bibnamefont {Kim}}, \ and\ \bibinfo {author} {\bibfnamefont
  {S.}~\bibnamefont {Cho}},\ }\href {\doibase 10.1038/ncomms13713} {\bibfield
  {journal} {\bibinfo  {journal} {Nat. Commun.}\ }\textbf {\bibinfo {volume}
  {7}},\ \bibinfo {pages} {13713} (\bibinfo {year} {2016})}\BibitemShut
  {NoStop}%
\bibitem [{\citenamefont {Chang}\ \emph {et~al.}(2018)\citenamefont {Chang},
  \citenamefont {Wu}, \citenamefont {He}, \citenamefont {Pei}, \citenamefont
  {Wu}, \citenamefont {Wu}, \citenamefont {Yu}, \citenamefont {Zhu},
  \citenamefont {Wang}, \citenamefont {Chen}, \citenamefont {Huang},
  \citenamefont {Li}, \citenamefont {He},\ and\ \citenamefont
  {Zhao}}]{chang_3d_2018}%
  \BibitemOpen
  \bibfield  {author} {\bibinfo {author} {\bibfnamefont {C.}~\bibnamefont
  {Chang}}, \bibinfo {author} {\bibfnamefont {M.}~\bibnamefont {Wu}}, \bibinfo
  {author} {\bibfnamefont {D.}~\bibnamefont {He}}, \bibinfo {author}
  {\bibfnamefont {Y.}~\bibnamefont {Pei}}, \bibinfo {author} {\bibfnamefont
  {C.-F.}\ \bibnamefont {Wu}}, \bibinfo {author} {\bibfnamefont
  {X.}~\bibnamefont {Wu}}, \bibinfo {author} {\bibfnamefont {H.}~\bibnamefont
  {Yu}}, \bibinfo {author} {\bibfnamefont {F.}~\bibnamefont {Zhu}}, \bibinfo
  {author} {\bibfnamefont {K.}~\bibnamefont {Wang}}, \bibinfo {author}
  {\bibfnamefont {Y.}~\bibnamefont {Chen}}, \bibinfo {author} {\bibfnamefont
  {L.}~\bibnamefont {Huang}}, \bibinfo {author} {\bibfnamefont {J.-F.}\
  \bibnamefont {Li}}, \bibinfo {author} {\bibfnamefont {J.}~\bibnamefont {He}},
  \ and\ \bibinfo {author} {\bibfnamefont {L.-D.}\ \bibnamefont {Zhao}},\
  }\href {\doibase 10.1126/science.aaq1479} {\bibfield  {journal} {\bibinfo
  {journal} {Science}\ }\textbf {\bibinfo {volume} {360}},\ \bibinfo {pages}
  {778} (\bibinfo {year} {2018})}\BibitemShut {NoStop}%
\bibitem [{\citenamefont {Pei}\ \emph {et~al.}(2011)\citenamefont {Pei},
  \citenamefont {Shi}, \citenamefont {LaLonde}, \citenamefont {Wang},
  \citenamefont {Chen},\ and\ \citenamefont {Snyder}}]{pei_convergence_2011}%
  \BibitemOpen
  \bibfield  {author} {\bibinfo {author} {\bibfnamefont {Y.}~\bibnamefont
  {Pei}}, \bibinfo {author} {\bibfnamefont {X.}~\bibnamefont {Shi}}, \bibinfo
  {author} {\bibfnamefont {A.}~\bibnamefont {LaLonde}}, \bibinfo {author}
  {\bibfnamefont {H.}~\bibnamefont {Wang}}, \bibinfo {author} {\bibfnamefont
  {L.}~\bibnamefont {Chen}}, \ and\ \bibinfo {author} {\bibfnamefont {G.~J.}\
  \bibnamefont {Snyder}},\ }\href {\doibase 10.1038/nature09996} {\bibfield
  {journal} {\bibinfo  {journal} {Nature}\ }\textbf {\bibinfo {volume} {473}},\
  \bibinfo {pages} {66} (\bibinfo {year} {2011})}\BibitemShut {NoStop}%
\bibitem [{\citenamefont {Liu}\ \emph {et~al.}(2012)\citenamefont {Liu},
  \citenamefont {Tan}, \citenamefont {Yin}, \citenamefont {Liu}, \citenamefont
  {Tang}, \citenamefont {Shi}, \citenamefont {Zhang},\ and\ \citenamefont
  {Uher}}]{liu_convergence_2012}%
  \BibitemOpen
  \bibfield  {author} {\bibinfo {author} {\bibfnamefont {W.}~\bibnamefont
  {Liu}}, \bibinfo {author} {\bibfnamefont {X.}~\bibnamefont {Tan}}, \bibinfo
  {author} {\bibfnamefont {K.}~\bibnamefont {Yin}}, \bibinfo {author}
  {\bibfnamefont {H.}~\bibnamefont {Liu}}, \bibinfo {author} {\bibfnamefont
  {X.}~\bibnamefont {Tang}}, \bibinfo {author} {\bibfnamefont {J.}~\bibnamefont
  {Shi}}, \bibinfo {author} {\bibfnamefont {Q.}~\bibnamefont {Zhang}}, \ and\
  \bibinfo {author} {\bibfnamefont {C.}~\bibnamefont {Uher}},\ }\href {\doibase
  10.1103/PhysRevLett.108.166601} {\bibfield  {journal} {\bibinfo  {journal}
  {Phys. Rev. Lett.}\ }\textbf {\bibinfo {volume} {108}},\ \bibinfo {pages}
  {166601} (\bibinfo {year} {2012})}\BibitemShut {NoStop}%
\bibitem [{\citenamefont {Tang}\ \emph {et~al.}(2015)\citenamefont {Tang},
  \citenamefont {Gibbs}, \citenamefont {Agapito}, \citenamefont {Li},
  \citenamefont {Kim}, \citenamefont {Nardelli}, \citenamefont {Curtarolo},\
  and\ \citenamefont {Snyder}}]{tang_convergence_2015}%
  \BibitemOpen
  \bibfield  {author} {\bibinfo {author} {\bibfnamefont {Y.}~\bibnamefont
  {Tang}}, \bibinfo {author} {\bibfnamefont {Z.~M.}\ \bibnamefont {Gibbs}},
  \bibinfo {author} {\bibfnamefont {L.~A.}\ \bibnamefont {Agapito}}, \bibinfo
  {author} {\bibfnamefont {G.}~\bibnamefont {Li}}, \bibinfo {author}
  {\bibfnamefont {H.-S.}\ \bibnamefont {Kim}}, \bibinfo {author} {\bibfnamefont
  {M.~B.}\ \bibnamefont {Nardelli}}, \bibinfo {author} {\bibfnamefont
  {S.}~\bibnamefont {Curtarolo}}, \ and\ \bibinfo {author} {\bibfnamefont
  {G.~J.}\ \bibnamefont {Snyder}},\ }\href {\doibase 10.1038/nmat4430}
  {\bibfield  {journal} {\bibinfo  {journal} {Nat. Mater.}\ }\textbf {\bibinfo
  {volume} {14}},\ \bibinfo {pages} {1223} (\bibinfo {year}
  {2015})}\BibitemShut {NoStop}%
\bibitem [{\citenamefont {Zeier}\ \emph {et~al.}(2016)\citenamefont {Zeier},
  \citenamefont {Zevalkink}, \citenamefont {Gibbs}, \citenamefont {Hautier},
  \citenamefont {Kanatzidis},\ and\ \citenamefont
  {Snyder}}]{zeier_thinking_2016}%
  \BibitemOpen
  \bibfield  {author} {\bibinfo {author} {\bibfnamefont {W.~G.}\ \bibnamefont
  {Zeier}}, \bibinfo {author} {\bibfnamefont {A.}~\bibnamefont {Zevalkink}},
  \bibinfo {author} {\bibfnamefont {Z.~M.}\ \bibnamefont {Gibbs}}, \bibinfo
  {author} {\bibfnamefont {G.}~\bibnamefont {Hautier}}, \bibinfo {author}
  {\bibfnamefont {M.~G.}\ \bibnamefont {Kanatzidis}}, \ and\ \bibinfo {author}
  {\bibfnamefont {G.~J.}\ \bibnamefont {Snyder}},\ }\href {\doibase
  10.1002/anie.201508381} {\bibfield  {journal} {\bibinfo  {journal} {Angew.
  Chem. Int. Ed.}\ }\textbf {\bibinfo {volume} {55}},\ \bibinfo {pages} {6826}
  (\bibinfo {year} {2016})}\BibitemShut {NoStop}%
\bibitem [{\citenamefont {Zeier}(2017)}]{zeier_new_2017}%
  \BibitemOpen
  \bibfield  {author} {\bibinfo {author} {\bibfnamefont {W.~G.}\ \bibnamefont
  {Zeier}},\ }\href {\doibase 10.1016/j.cogsc.2017.02.003} {\bibfield
  {journal} {\bibinfo  {journal} {Curr. Opin. Green Sustain. Chem.}\ }\textbf
  {\bibinfo {volume} {4}},\ \bibinfo {pages} {23} (\bibinfo {year}
  {2017})}\BibitemShut {NoStop}%
\bibitem [{\citenamefont {Ashcroft}\ and\ \citenamefont
  {Mermin}(1976)}]{ashcroft_mermin_1976}%
  \BibitemOpen
  \bibfield  {author} {\bibinfo {author} {\bibfnamefont {N.~W.}\ \bibnamefont
  {Ashcroft}}\ and\ \bibinfo {author} {\bibfnamefont {N.~D.}\ \bibnamefont
  {Mermin}},\ }\href@noop {} {\emph {\bibinfo {title} {Solid State Physics}}}\
  (\bibinfo  {publisher} {Saunders College Publishing},\ \bibinfo {year}
  {1976})\BibitemShut {NoStop}%
\bibitem [{\citenamefont {Yu}\ and\ \citenamefont
  {Cardona}(1996)}]{yu_cardona_1996}%
  \BibitemOpen
  \bibfield  {author} {\bibinfo {author} {\bibfnamefont {P.~Y.}\ \bibnamefont
  {Yu}}\ and\ \bibinfo {author} {\bibfnamefont {M.}~\bibnamefont {Cardona}},\
  }\href@noop {} {\emph {\bibinfo {title} {Fundamentals of Semiconductors:
  Physics and Material Properties}}}\ (\bibinfo  {publisher}
  {Springer-Verlag},\ \bibinfo {year} {1996})\BibitemShut {NoStop}%
\end{thebibliography}%

\end{document}